# APPLICATION OF THE EARTH'S NATURAL ELECTROMAGNETIC NOISE TO GEOPHYSICAL PROSPECTING AND SERACHING FOR OIL.

Sergey Yu. Malyshkov [1, 2)], Yury P. Malyshkov[1, 2)], Vasily F. Gordeev [1, 2)],
Sergey G. Shtalin [1, 2)], Vitaly I. Polivach [1, 2)], Yury Yu. Bazhanov [2)], Terje Hauan[3)]

[1] Institute of Monitoring of Climate and Ecosystems,
Siberian Branch of the Russian Academy of Science, Tomsk, Russia
[2] «Emission» Ltd., Tomsk, Russia
[3] Emtek Monitoring, Stavanger, Norway

*When applying the Earth's natural pulse electromagnetic fields to geophysical prospecting one should take into account characteristics of their spatial and temporal variations. ENPEMF is known to include both pulses attributed to atmospheric thunderstorms and pulses generated in the lithosphere by mechanic-to-electric energy conversion in rocks. It is evident that the most valuable information on the geophysical structure of a certain area is obviously contained in pulses originated from this area*

*This article covers a method of recording spatial variations of the Earth's natural pulse electromagnetic fields which is able to take due account of spatial and temporal variations of EM fields and suits to reveal crustal structural and lithologic heterogeneities including hydrocarbon pools.*

*We use a system of several stations recording the ENPEMF concurrently to erase the temporal variations from ENPEMF records and to sort out the pulses of local and remote origin. Some stations are fixed (reference) and record only temporal variations of EM fields. While the other stations are mobile and measure pulse characteristics related to both spatial and temporal ENPEMF variations along measurement routes crossing the area investigated. Spatial variations of EM fields left after having deleted the temporal variations and pulses generated out of the area investigate show the availability or the lack of geophysical anomalies.*

## Introduction

The term "Earth's natural pulse electromagnetic fields" or ENPEMF was introduce by A.A. Vorobyov in the late 60ties of the last century. It was him who expressed a hypothesis that pulses can arise not only in the atmosphere but also within the Earth's crust due to processes of tectonic-to-electric energy conversion (Vorobyov, 1970; Vorobyov, 1979). An intensity of pulse flux was expected to be increased in the eve of and at the moment of large earthquakes.

This hypothesis was being actively developed in the end of the last century (Electromagnetic, 1982; Gokhberg et al., 1985; Gokhberg et al., 1988; Surkov, 2000). But at present the number of publications on the topic keeps going down because expectations for higher accuracy of application of NPEMFE to earthquake prediction have not been confirmed.

Approximately at the same time with searching for and investigating the earthquake precursors, a method of recording the Earth's natural electromagnetic noises started to be applied to geophysical prospecting. The method was mainly used to study land sliding processes (Salomatin et al., 1981; Mastov et al., 1984) because, with the frequency of a few kilohertz, one considered signals to be recorded only at m-scale depth and not deeper. And the NPEMFE method got neither a wide acceptance in nor an extensive application to geologic engineering. A flux of pulses recorded was of a noise character and irregular pattern. Even a firmly-fixed recording unit could record about a hundred pulses within the first second, but a second later there could be no pulses recorded at all. Such recorded pulses included both pulses of atmospheric origin, so called "atmospherics", arisen due to constantly occurring thunder storms, and pulses generated by crustal rocks and arisen due to yearlier-unknown mechanisms. The availability of pulses of atmospheric, lithospheric and anthropogenic origin in a recorded pulse flux resulted in bad reproducibility of results.

There have been made many attempts using various ways and methods to enhance the quality of geophysical data. Usually one increased the sensitivity of recording stations so much that the number of pulses recorded significantly exceeds the number of pulses of atmospheric origin. About a hundred of lightning discharges per second is thought to occur on the globe; therefore recording units were tuned for such sensitivity that the unit records more than 100 pulses per second. However such a way is hardly able to improve the data reliability. The point is that pulses generated by lightning discharges can penetrate into the ionosphere and magnetosphere and produce a noise component. And the number of noise pulses increases exponentially as their amplitudes get higher. Thus, instead of expecting positive results, one can completely miss the information on the subsurface geologic structure.

The presence of clear diurnal variations makes the task of application of the ENPEMF method to geophysical prospecting dramatically complicated. Regardless a noise character of the pulse, there are six- and eight-hour, semidiurnal, diurnal and even semiannual and annual periodicities in spectral characteristics of ENPEMF (Malyshkov, Malyshkov, 2009). Having so many challenges and such influencing factors, acquiring the trustworthy and accurate geophysical data would seem to be impossible. And the fact that there are only few publications on application of the ENPEMF method to geophysical survey just confirms this conclusion.

However, to our opinion, the main reason for the failure of ENPEMF application to both the earthquake prediction and geophysical prospecting is in the fundamental physical background of the proposed method.

In (Malyshkov, Malyshkov, 2009) we have assumed and have given evidences in favor of the hypothesis that it is deep-seated strain waves that are main sources of a noise component of the Earth's natural pulse electromagnetic field. The strain waves appear due to eccentric motion of the Earth's core and lithosphere. Traveling from the mantle up to the surface such waves generate mechanic-to-electric energy conversion in rocks and, thus, produce a pulse flux recorded by recording stations. This physical concept about constantly acting near-surface lithospheric sources of the EM field was accepted as a basis for the geophysical prospecting methods mentioned below.

Thus, when applying the Earth's natural pulse electro magnetic fields to geophysical prospecting the following should be taken into account:
1. The pulses recorded are of a noise character and of irregular pattern in time. The number of pulses recorded goes exponentially up as the discrimination limit of recording stations is getting lower.
2. There is a clear diurnal ENPEMF variation and such variations are of irregular pattern and vary greatly within a year.
3. Spectral characteristics of temporal variations include a great number of split bands.
4. The flux of pulses recorded includes pulses of lithospheric and atmospheric origin, as well as strong pulses generated by remote sources. It also includes technogenic pulses.
5. Most pulses recoded are from sources located out of the research area.

It should be noted that none of the existing geophysical methods is able to take such characteristic features of ENPEMF into consideration in full. Most commonly the fact that pulses mostly arise in sources located out of the interesting area is neglected. Our years-long measurement have shown that up to (80-90) % of pulses recorded by measuring stations are generated by remote sources located out of the area investigated. Such pulses carry no information on the geologic structure of the area. Time instability of the ENPEMF, amplitudes and phases of the signal are often improperly interpreted as fields' spatial variations related to geophysical heterogeneities, whereas they are just field temporal variations. Thus the proposed methods can unlikely be used for geophysical prospecting of deep-seated objects including oil and gas fields. Because anomalies directly related to oil and gas pools are weak and disguised with much stronger temporal variations of ENPEMF from remote sources located out of the oil and gas field.

It is obvious that pulses directly arisen at a given point of the area are able to give the most valuable information on its geophysical structure.

### Ways of getting spatial variations of EM noise

We suggest removing the pulses arisen due to atmospheric processes and pulses generated out of the interesting area from the flux of pulses recorded. The signal is "cleaned" from irrelevant pulses when data are being recorded and processed. This can be done by:

- applying a wide-area network of mobile and fixed EM noise recording stations;
- tuning the recording stations to optimal and approximately same sensitivity and same discrimination limits and by adjusting all the stations identically;
- by sorting out pulses produced by remote sources from pulses of local origin.

Thus, the flux of pulses recorded is defined with both spatial and temporal variations of EM fields. When conducting geophysical prospecting temporal variations of EM fields and all the pulses arisen from remote sources should be erased from the recorded signal. As it has already been mentioned this can be done with the help of several stations recording the ENPEMF concurrently. Some recording stations are fixed and serve as reference ones; they record only temporal variations of the ENPEMF. The others are mobile units and record both temporal and spatial variations along routes crossing the area investigated. Our method applies no less than two recording stations and the accuracy with which anomalies can be revealed increases with the increase in the number of recording stations used.

Pulses of local origin can be distinguished from remote ones by time of arrival and amplitude of pulses recorded by a network of wide-spaced stations. Pulses produced by remote sources, for example atmospherics, will propagate along the Earth-ionosphere waveguide and reach recording stations located at a small distance from each other at approximately the same time; and they will have the same amplitudes. Pulse signals arisen due to large lithospheric objects will reach the surface and will further travel as a ground ray as fast as the light, and will damp only slightly. Therefore all the recording stations will record such pulses concurrently and of about the same amplitudes.

One will observe the different picture for pulses of local lithospheric origin, i.e. pulses arisen in the crust at a small distance from recording stations. Such pulses will travel to recording stations mostly through rocks. Heavy damping of EM fields in the Earth' crust will result in significant difference in amplitudes of pulses recorded directly above the signal source and pulses arisen at a distance from it. When using recording stations of discrimination limit behavior which do not record low-amplitude pulses, there can appear a situation when a more remote recording station will record less pulses per a certain period of time than the station located directly above the emitting geophysical anomaly. But if single pulses originated from a local source have rather high amplitudes and are recorded by all the spaced-apart stations, the amplitude of pulses recorded will significantly vary depending on the distance between the signal source and the recording station. Particularly this phenomenon is taken as a basis for the stations and method we have developed.

**Recording stations: tuning to optimal parameters**

Most measurements were carried out with multichannel recording stations MGR-01. The MGR-01 stations were designed, developed and produced by this article's authors in the Institute of Monitoring of Climate and Ecosystems, a Siberian Branch of the Russian Academy of Science. The stations are designed for permanent monitoring of ENPEMF characteristics and geodynamic processes occurring in the Earth's crust, and for filed geophysical measurements. The MGR-01 design allows their unstaffed operation and data communication via radio or cellular phones. The MGR-01 stations are certified (certificate No 24184), registered in the State register of Measuring Tools under the reference number 31892-06, and allowed for use in the territory of the Russian Federation.

The stations record a magnetic field component being received in a narrow very low frequency (VLF) band by two antennas in two orthogonal directions (N-S and W-E). Note that the stations receive only the signals of certain frequencies that are above a user-specified limit (discrimination limit) and can measure fields from $2 \cdot 10^{7}$ to 400 A/m or from $2.5 \cdot 10^{-4}$ nT to $5 \cdot 10^{-4}$ T.

According to our many-year experience, very high accuracy of instruments is a necessary prerequisite of reliable geophysical data. And ENPEMF temporal variations in the VLF band are governed by diurnal and annual rhythms of crustal movement, the diurnal rhythm depending on the date, geographical references of the area and its geophysical properties.

When the instrument sensitivity is too high, the flux of pulses predominantly contains noise components of atmospherics and interference pulses. In case the instrument sensitivity is low, the stations receive only pulses arisen due to large thunderstorm discharges and do not record any pulses originated form the local area. Therefore, stations sensitivity should be optimal. We conducted our measurements at a discrimination limit of 10 nT, and a resonance frequency of 14-17 kHz.

To tune the stations to optimal sensitivity, special calibration relations have been used. These calibration relations were developed in the course of our many-year research of the Earth's natural pulse electromagnetic field in various regions of the Eurasia (Malyshkov and Malyshkov, 2009). When tuning the stations, waveforms of temporal variations recorded by stations and a diurnal pattern typical of a given season are to be same. When the stations have been tuned to optimal sensitivity, it is necessary to have all the stations similar (identical). Such similarity adjustment is very important and needs to be done very thoroughly and carefully. It is the similarity of fixed and mobile stations that defines the reliability and trustworthy of geophysical data. In case the stations are not similar, different stations will receive the same pulse from the same source differently. This will result in errors when distinguishing pulses of local origin from pulses arisen from remote emitters, and, as a consequence, will cause the diminution of the method accuracy. When adjusting the stations identically, all the stations were placed next to each other and their antennas were oriented in a required direction. A maximum similarity in readings of all the stations is achieved by adjusting parameters of stations' measurement channels. Stations sensitivity similarity was verified by both the number of pulses recorded by the stations per unit time and time of arrivals of single pulses.

Fig. 1, *a* illustrates records from two stations. Both stations record signals by their W-E antennas for random 250 sec. The data from Station 0A are multiplied by a factor of – 1 for better imaging. One can see that the two stations record signals at the same moments of time and the number of pulses per unit time varies very slightly.

This slight difference in records from different stations was eliminated by applying corrections. The corrections were taken from an earlier-drawn correction graphs (Fig. 1, *b*).

The correction graphs are absolutely necessary because it is impossible to have receiving antennas, filter characteristics, amplifiers and other station's components absolutely identical.

To draw out the correction graphs, all the stations had been concurrently recording temporal variations of ENPEMF for 24 hours or for certain several hours specified before the measurements to the intent that thereafter to carry out field measurements at the same certain hours.

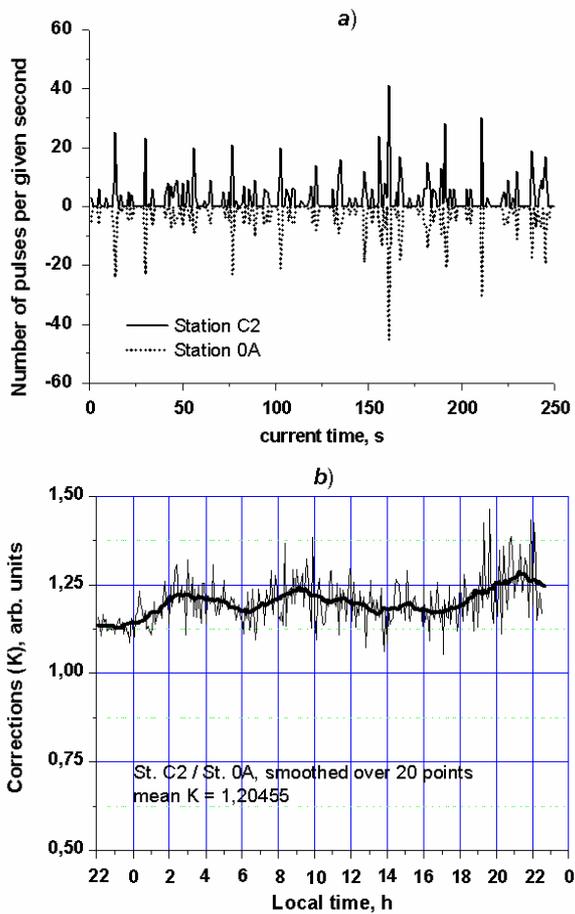

**Fig. 1. Example of verifying the two stations identity and constructing a correction graphs.**

Then the graphs of temporal variations obtained were compared to each other in order to select a "reference" station the records from which had the nearest value to the measurement average from all the stations. Finally correction graphs (similar to Fig. 1, *b*) were developed for each station and each receiving channel. The graphs represent the difference between records from a certain station and records from the reference one.

When conducting field measurements, first fixed stations were continuously recording the number of pulses arrived for a time unit (usually 1 s) at a certain point of the area. Then mobile stations were placed at a first measurement point. They were measuring for a time unit specified before the measurements (usually 3-5 min.) with the same record interval (1 s). When measurements on the first point were completed, the mobile stations were moved to next one and the procedure was repeated. When the profile measurements were completed, we performed the statistical processing of the data obtained.

The data processing procedure was as follows.

We found mean intensity for a certain mobile station and a certain receiving channel, and mean intensity of a signal recorded at the same time by a similar channel of one of fixed stations at a given measurement point. Then, applying correction graphs, we corrected the difference in records from these stations relative to the reference station.

Then we calculated the differential of records from this mobile station on the given point relative to the fixed station by subtracting the fixed station records from the records of the mobile station; or by dividing the mobile station records by the fixed station records. After having processed the data from all the measurement points, we obtained two-dimensional route variations of records from the given station relative to the records from given fixed station. The same procedure was applied to find route variations in ENPEMF intensity but by the difference between the every other mobile and fixed stations. The mean intensity was found for each pulse flux arrived at each receiving channel of each station.

Basing on the above-mentioned data we constructed ultimate ENPEMF intensity variations. Correction of each station data for data from the fixed (reference) station provides more reliable ultimate results.

And spatial variations of fields were also analyzed at each receiving channel for a given route. Then we made a conclusion if a geophysical anomaly is available, mapped its boundaries and interpreted geologic data.

## Examples of applying the ENPEMF method to geophysical prospecting

### a) Two-dimensional survey

The reproducibility of geophysical data obtained with the ENPEMF method was proved along a survey line (a measurement route) though the Urbinsky thrust. This thrust is the most significant tectonic disturbance near the city of Tomsk. Measurements were carried out along a 2km-long route in different years, seasons and under various weather conditions. Some pieces of the measurement were plotted in Fig.2. In different measurements there were used different methods to delete temporal variations from records. Therefore the curves may be compared only qualitatively. Fig. 2 illustrates the sharp decrease in ENPEMF intensity that reveals an EM field anomaly at the W-E channel on Measurement point 12. Fine tuning and high identity of fixed and mobile stations ensured clear delineation of the anomaly in different days and in different seasons even in winter when snow laid thick upon the ground. Note in most cases "useful", i.e. local, pulses amount 20-30% of the number of total pulses recorded at this or that measurement point. It means that about 80% of pulses are recorded concurrently by both fixed and mobile stations. Hence, pulse sources were far beyond the limits of the area investigated. Similar results were obtained even in thunder stormy days when during measurements we saw flashes of lightning across the sky and heard growls of thunder. When fixed and mobile stations are tuned finely they record atmospherics and signals arisen from remote pulse sources concurrently and very similarly. Therefore such signals are easily deleted from records when defining spatial ENPEMF variations.

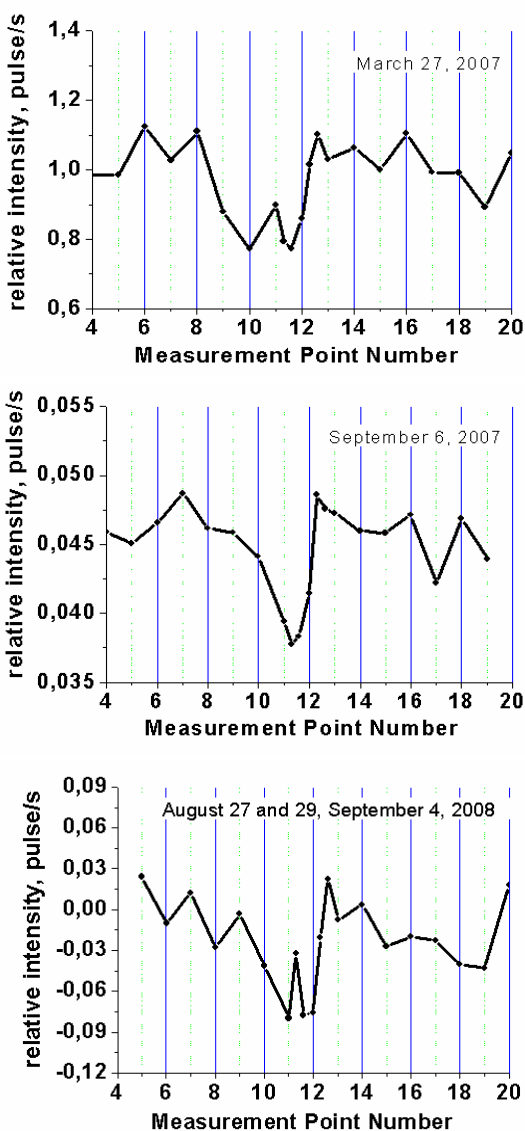

**Fig. 2. Data reproducibility on the Urbinsky thrust.**

The geophysical anomaly in Fig.2 is most likely related to one of the faults "feathering" the Urbinsky thrust. On the terrain the anomaly is confined to a long ravine framed in a gentle slope on the one side and in a high bank on the other side.

Now let us have a look at an example of revealing a fault in the Krasnoyarsk region. A measurement route crossed a fault of indistinct morphology and travel time characteristics. The fault was revealed on the basis of aero- and satellite image interpretation. Measurement points were at a distance of 50 meters from each other. Measurements were taken in 1 s for 5 minutes on each measurement point. Therefore not less than 300 measurements of ENPEMF intensity per measurement point were totally done at each N-S and W-E receiving channels. Fig.3 illustrates ENPEMF intensity data. The most significant anomaly was revealed at the both N-S and W-E channels between Measurement points 65-95.

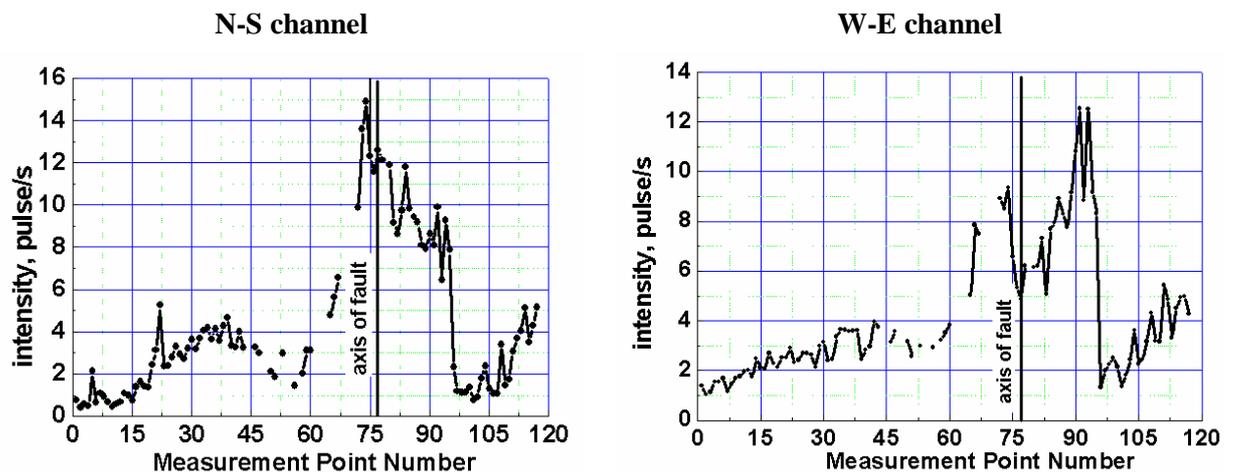

**Fig. 3. Variation in ENPEMF intensity along the route crossing a geologic fault**

It was there that the route crossed an axial region of the fault dividing neo-tectonic blocks.

Fig. 4 shows an example of application of equipment and the method developed to analysis of natural climatic systems. A right bank of the river Ushaika near Tomsk served as a model of such a system. According to geologic data the Ushaika's watercourse is confined to a geologic fault. The river bank was like a steep slope but slightly flattened in the middle of the measuring route. The forest on the flattened area was highly damaged by thunderstorm activities. Many trees had their bark and tops burnt. Measurements were carried out with three mobile and two fixed stations moving down and then up the slope on September 16, 2008. The measurements were taken in each 25 m. Each station recorded signals at both N-S and W-E channels. In Fig.4, *a* one can see the decrease in EM noise at both receiving channels (N-S (Fig. 4, *a*) and W-E (Fig.4, *b*)) as the axial region of the fault is getting closer.

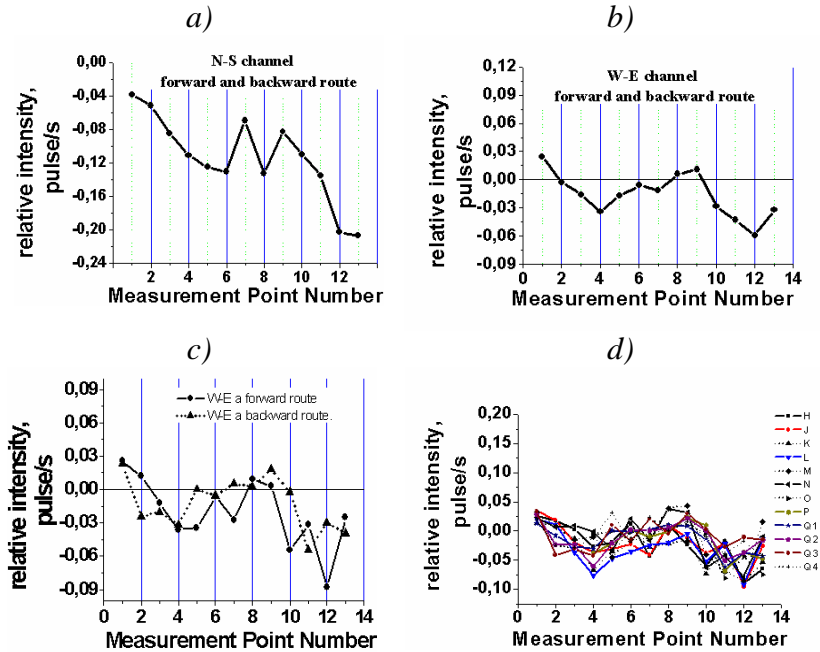

**Fig. 4. Variation in low-frequency radio noise of the Earth along a route crossing a geologic fault and thunderstorm-damaged area**

Mean values (after having deducted the records of fixed station from mobile station records) of EM noise intensity recorded in two-way directions (down and up the slope) are plotted on the ordinate axis. In Fig.4, *c* one can see good data reproducibility in both measurement directions. The good data reproducibility is most clearly seen in Fig. 4, *d,* which illustrates all the 12 measurements recorded at the W-E channel in both directions (6 forwards and 6 backwards) by different MGR stations. One can clearly see a thunderstorm-damaged area between Measurement Points 5-10.

*b) Areal measurements, defining the oil and gas field boundary and productivity*

Abilities of the ENPEMF method can be demonstrated with areal measurements of EM noises recorded by widely-spread stations. Let us show some examples of such areal measurements.

First the areal measurements were carried out on an existing deposit of Lithium in Western Finland from June 28 till July 9, 2009. There were set two tasks; i.e. to check if the method is able to work and to rectify the boundary of lithiumiferous rocks. Before the work commenced we were informed about a 100*400 m$^2$ and 200-deep pegmatite intrusion. Fig. 5, *a,* shows an intrusion modeling based on drilling data and presented by the Keliber Resources Ltd Oy.

Measurements were carried out totally with 4 mobile and 4 fixed recording stations along two routes simultaneously. Two of the 4 fixed stations were placed in the deposit territory and the other two stations–out of it. Therefore two stations recorded at each measurement points. Fig. 5, *b,* illustrates measurement routes and measurement points. Where the intrusion cropped out to the day, the distance between measurement points was 25-50 m. But as the intrusion limits were getting farther the distance between points was gradually increased up to 100 m. After having processed the data obtained and calculated the spatial ENPEMF variations there was constructed a map of ENPEMF anomalies. Then the map constructed was superimposed upon an existing deposit map (Fig. 5, *b*). One can see the lithium-containing territories by lower ENPEMF intensity.

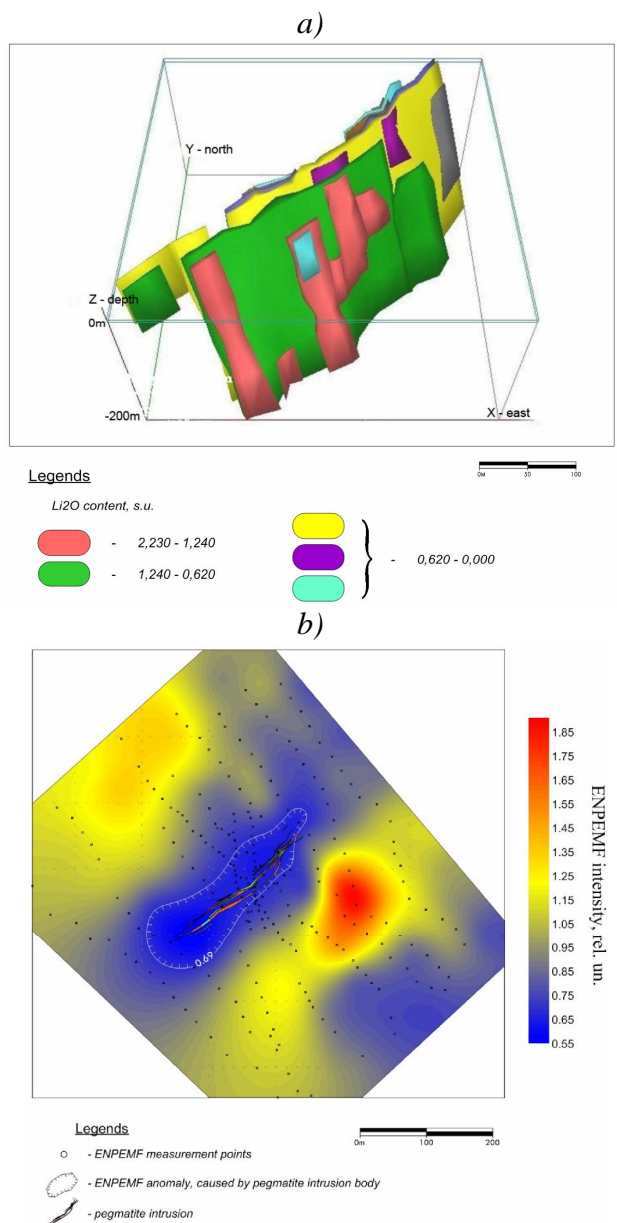

**Fig. 5. ENPEMF data recorded on lithium deposit in Finland.**

Moreover the pegmatite intrusion is also within the ENPEMF anomaly boundary line. Judging by the results the lithium deposit boundaries are wider than one thought before the measurements. A south-eastern part of the deposit might also be prospective. But exploratory wells were not drilled here to prove.

Application of natural electromagnetic noises of the Earth to searching for hydrocarbon fields is based on the fact that many mineral deposits including hydrocarbon fields are confined to zones of higher crustal heterogeneity, to geologic faults and their intersections. Researches conducted in the Tomsk and Krasnoyarsk regions, in Tatarstan and Udmurtia have proved that a hydrocarbon prospect is shown up with a certain emitting zone around it and a "silent" interior zone. This "silent" interior zone, i.e. a zone of lower EM field intensity, is located directly above the hydrocarbon pool. These features prove that geologic structures located on a productive area are very slow-moving and either there are no faults and cracks there or they are "hermetically sealed". Particularly it is such structures that ensure hydrocarbons to be accumulated.

Also our ENPEMF method and equipment were used to define boundaries of two oil fields in Udmurtia. The work was conducted in November 2008. Totally twenty MGR-01 stations (ten fixed and ten mobile stations) recorded the Earth's natural pulse electromagnetic field concurrently. The fixed stations were placed in a non-productive part of the oil fields.

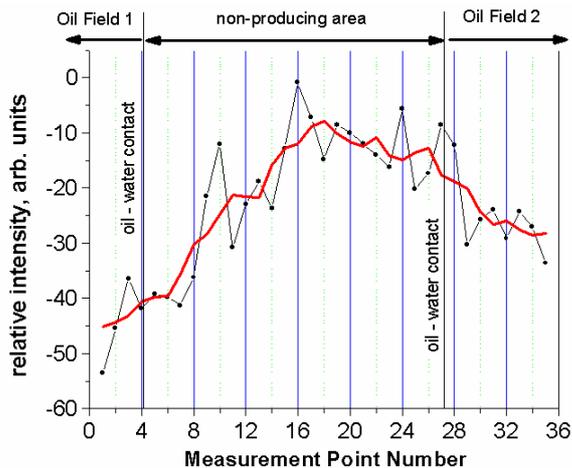

**Fig. 6. Variation in pulse flux intensity along a route crossing two oil fields**

Data of measurements are plotted in Fig.6. One can see that the signal intensity, compared to that of non-productive parts, becomes lower as productive parts of Oil Fields 1 and 2 are getting closer. Consequently it is possible to apply the ENPEMF method not only to defining the oil field boundaries but also to assessing the productivity of field's selected areas. Depth of reduction in the ENPEMF intensity can serve as a measure of area productivity. In this case the total net pay thickness of reservoirs was 15-18 m for Field 1 (in the beginning of the measurement route) and about 10 m for Field 2 (in the end of the route). Oil-water contact is shown in Fig. 6 with vertical blue bars.

The central area of Field 2 was investigated in more details. The Field 2 looked like two close anticlines separated by a non-productive area (Fig.7). After having conducted the areal measurements (measurement routes and points are shown in Fig. 7, *a*), processed the data and constructed the ENPEMF anomaly map, the map was superimposed upon a subsurface structure map of reservoir top to analyze them visually and to interpret the data further. Spatial variations of ENPEMF recorded at the N-S channel are shown as spots of various colors corresponding to certain intensity of EM noises. The color also illustrates the relation of signal intensity in a given point to the intensity of the signal recorded at the same moment of time in a given reference point on the area. The data obtained were normalized from 0 to 1 and smoothed. To make the measurement more precise and accurate pulses were recorded concurrently both at fixed (reference) and measurement points with no less than 7-8 MGR-01 stations. One can find detailed information on MGR-01 recording stations, the measurement procedure and ways of data processing in (Malyshkov et al., 2009, *b*).

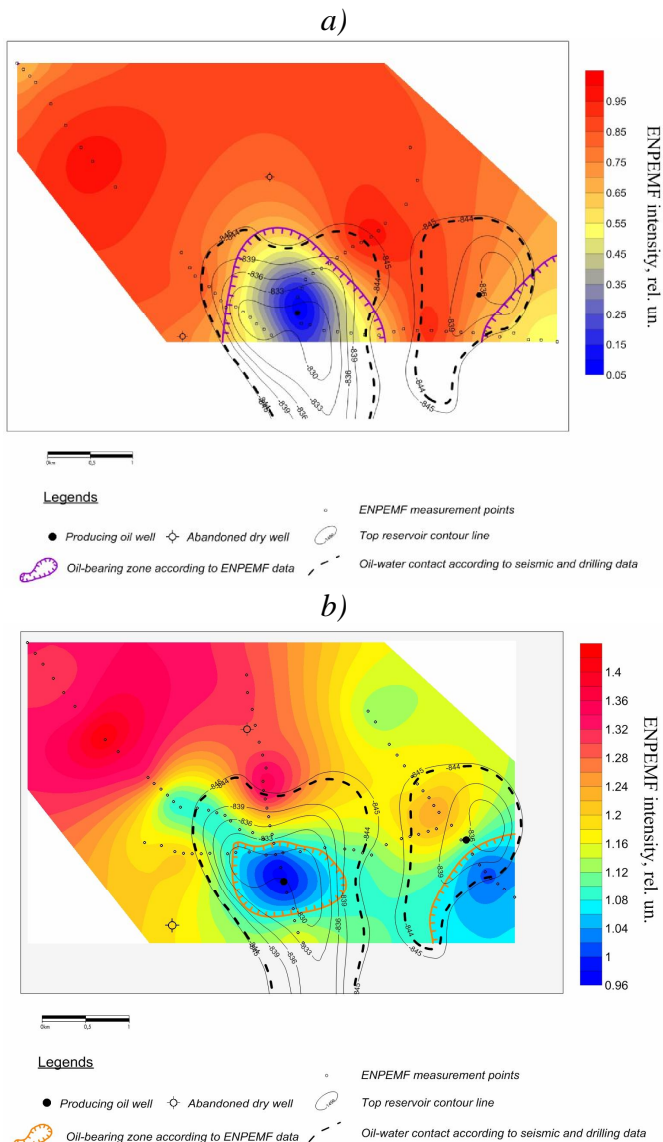

**Fig.7. ENPEMF data recorded in an oil field in Udmurtia**

A year later (late November – early December 2009) similar measurements were conducted in the same field that made it possible to assess the reproducibility of the data obtained earlier (Fig.7, *b*). Note that in 2008 fixed stations serving as reference ones (ENPEMF intensity variations were assessed with the reference to fixed stations) were placed at other points on productive and non-productive areas of the oil fields than they were placed in 2009. Layout charts and ways of data processing were also different in 2008 and 2009. Therefore Figs. 7, *a* and 7, *b* can be compared only in a quality manner.

However one can clearly see that data obtained in different years agree with each other. Thus the proposed ENPEMF method has proved good data reproducibility. It should be noted that in both cases the results obtained completely coincide with the oil-bearing areal limits (an isoline with TVDSS of - 844 m). Slight discrepancies in a right side of Fig. 7 are likely caused by a small reservoir's net pay thickness (it is hardly 2-m thick) and insignificant number of measurement points on this area of the field. Note that a trap on the left was more than 10 m net pay thick. It also should be noted that a relative intensity of spatial ENPEMF variations is minimal above most productive areas of the oil field.

In conclusion of this chapter let us compare the given geophysical method with other conventional methods as exemplified by an oil field in Tatarstan. During this work we tried to solve a task of searching for new prospects adjacent to existing productive areas of the field.

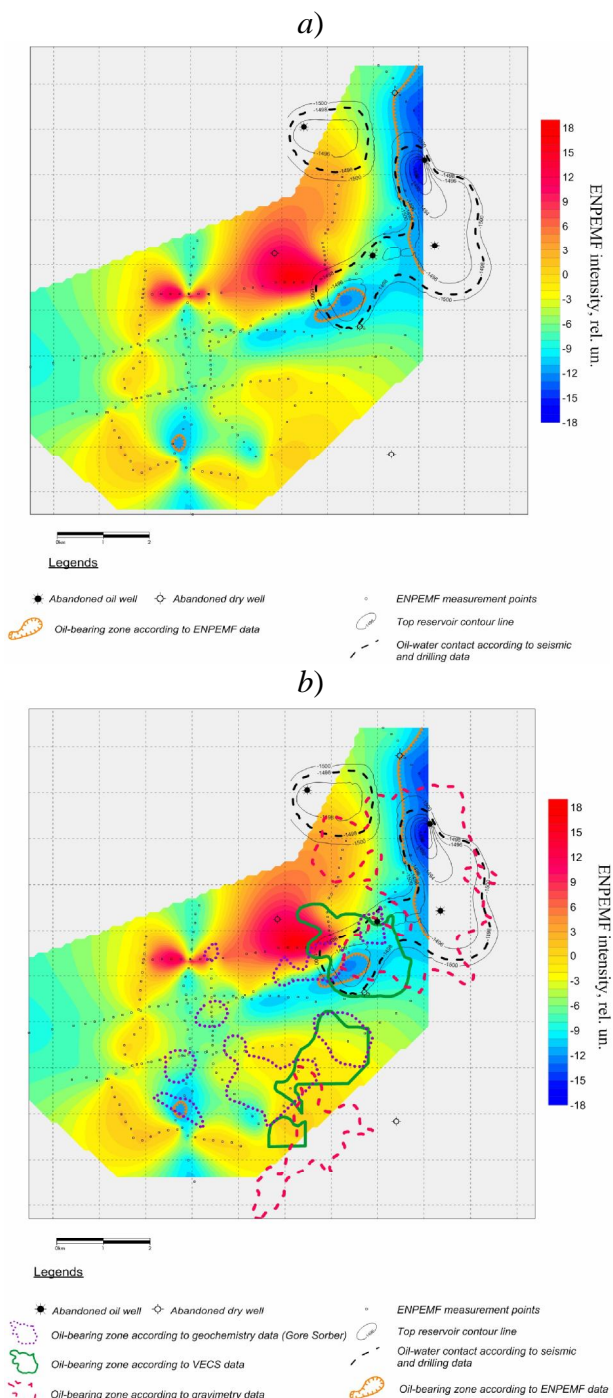

**Fig. 8. ENPEMF data from oil pool in Tatarstan**

Fig. 8, *a* shows a layout chart of measurement routes and points. The measurements were carried out in February 2010. Data of EM noises recorded were processed and presented as colorful spots. ENPEMF measurement points were at a distance of 200 m from each other. Here in Fig.8 one can see oil-water contacts revealed by drilling and seismic data before our measurements.

It is also seen that hydrocarbon prospects revealed by our ENPEMF method and shown as dark blue spots are in a north-eastern part of the measurement area. ENPEMF anomalies exactly coincide with areal limits of oil-bearing formation of 2 m thick and thicker. Judging by our measurements, areas adjacent to the field are less hydrocarbon-promising.

Let us compare our data with data obtained by other geophysical methods including vertical electric circuit sounding

(VECS), geochemical and gravity surveys (Fig. 8, *b*). Such data on geology, drilling, VECS, geochemical and gravity surveys were kindly provided at our disposal by TNG-Kazangeophysica Ltd but only after having finished the field work and having distributed the ENPEMF measurement results to the Customer.

As one can see the ENPEMF data exactly coincide with seismic acquisition and drilling data. In a north-eastern part of the work area the ENPEMF anomaly coincides well with oil-bearing limits according to gravity survey and coincides, to some extent, with limits according to VECS and geochemical survey data. This complex analysis makes it possible to conclude that planning of development drilling is more accurate according to ENPEMF data and, in comparison with other low-cost geophysical methods, the ENPEMF method is much less risky of drilling a dry well.

**Most probable mechanism of spatial and temporal ENPEMF variations**

According to classical thermodynamics electromagnetic waves in a kilohertz range cannot travel up from the depth of several hundred meters and deeper due to the skin-effect and strong absorption of signal by rocks. Therefore ENPEMF anomalies appearing above hydrocarbon fields seated at the depth down to 3 km can hardly be related to signal generated by the formation itself. The reason for such anomalies is more likely surface sources somehow related to the subsurface crustal structure. Several reasons for this interrelation can be assumed.

The electric topography of solids' surface is known to reflect the internal structure of and structural imperfections in solids. Contact potential difference is developed on the phase-phase interface in the multiphase geologic media (Surkov, 2000). There might also appear double electric layers having the different charge density in the hydrocarbons-water-saturated rocks interface above the oil pool and in its vicinity due to various electro kinetic processes. The topography of charged area distribution in near-surface rock layers can reflect the electric structure of deeper geologic layers and cause ENPEMF anomalies.

ENPEMF anomalies can also be caused by property differences in rocks above hydrocarbon fields. Hydrocarbons migrating from the oil pool cause formation of magnetic minerals, sharp increase in magnetic property dispersion and changes in rock density and P-wave velocity (Berezkin et al., 1978). Spatial distribution of these parameters is driven by location of oil and gas accumulations and cause formation of peculiar slightly-altered rocks above oil pools.

At the same time anomalies well agree topographically with oil-bearing areas and pool boundaries.

Seismic-to-electric energy conversion in rocks can also be a reason for abnormal behavior of EM fields. In recent years there have been recorded anomalies of micro seismic noises above oil fields (Goloshubin et al., 2006). Conversion of energy of microseismic vibrations being recorded even on the Earth's surface to electromagnetic pulses can also make oil pools be shown up as ENPEMF anomalies.

However, according to our invincible belief, ENPEMF anomalies appearing above structural and crustal heterogeneities including even ones at a km-scale depth can most likely be caused by subsurface deformation waves. We have already mentioned our hypothesis stating that eccentric motion of the Earth's inner core and the lithosphere can give rise to zones of high and low pressure in the fluid surrounding the Earth's core. According to the hypothesis (Malyshkov, Malyshkov, 2009), elastic strain from such zones transfers through the mantle up to the Earth's surface in a few minutes. The diurnal rotation of the Earth produces waves on its surface which like a tidal wave spiral the Earth from east to west. As the Earth rotates, points on its surface move relative to the less mobile zones of high and low pressure in the fluid core approaching them at 300 m/s (1000 km/h) or faster in the middle latitudes. Such a high velocity of moving through the mantle and lithosphere extension zones loosens traction between crustal structural elements, intensifies processes of rock cracking and crack opening, and activates development of double electric layers, tribo – and electro – kinetic processes and movement of water and other fluids. This will result in recording a higher intensity pulse flux from the lithosphere by the MGR-01 stations. Then a compression wave follows an extension wave and the process retards generation of EM pulses.

In our opinion particularly such extension and compression waves govern a periodicity in the Earth's natural pulse electromagnetic and other geophysical fields. Moreover they produce annual, diurnal, semidiurnal, 8- and 6-hour periodicities.

If such deformation waves really exist then, when moving from the lower mantle up to the Earth's surface, they will travel through various geologic structures, interact with them and change their own characteristics. Waves coming up against oil- and gas-bearing formations are to be different by their characteristics from waves that did not meet any of such obstacles. Thus the difference in ENPEMF intensity we observed can be related to not different concentration of surface pulse sources but to different paths of propagation of subsurface srain waves triggering such pulse sources.

If we are right in our interpretation, there appear unique possibilities to apply subsurface strain (deformation) waves to solving fundamental and applied tasks. Propagating from the lower mantle the strain waves carry information on the core motion (almost in real time) but also, like X-rays, on the entire path of their propagation from the incredible depth up the surface. A near-surface crustal layer trigged by strain waves generates pulses. The pulse flux and differential of pulse fluxes contain the information on difference in paths of deformation wave propagation and on availability of geophysical anomalies including oil and gas ones along the propagation paths.

Having in mind such a mechanism of triggering near-surface EM pulse sources it is necessary to take into account effects related to climbing extension and compression waves from east to west and from the lower mantle up to the surface (Malyshkov, Malyshkov, 2009). Structural heterogeneities including faults, cracks and lithologic interfaces will respond independently and "work" as single pulses recorded at different moments of time. It is likely for this reason the pulses recorded at N-S channels often appear to disagree by the arrival time with pulses recorded at W-E channels. However both orthogonal directions of receiving pulses (N-S and E-W channels) are modulated by low-frequency strain waves produced by eccentric motion of the Earth's core.

We consider this as a reason for why, unlike the rather long time series show high correlation in N-S and E-W data, time of pulse arrival at a N-S channel is slightly correlated with arrival time at a E-W channel. These peculiarities in pulse signals receiving have been mentioned in our earlier publications.

A path of core motion inside the Earth has been revealed and discussed in (Malyshkov, Malyshkov, 2009) using the mechanism mentioned above and many-year ENPEMF observation. It has been shown that the inner core is never at the Earth's geometric center but it oscillates near the center along a closed orbit. The plane of the core orbit is normal to the equatorial plane and tilted 45° to the direction to the Sun and to the Earth's orbit. It is the path of annual motion of the core relative to the planet geometric center that governs seasonal variation of ENPEMF diurnal patterns, Earth's seismicity and other geophysical fields.

Over recent months the article's first two authors managed to discover a number of additional sound arguments to prove the existence of strain waves produced by the Earth's core.

Thus the spectra obtained from the spectral analysis in (Malyshkov, Malyshkov, 2009) surprisingly lack the lunar component in spectral bands of EPEMF and seismicity. There appears an impression that the Earth's core does not react to the Moon's gravitational attraction. To explain such a paradox, EPEMF spectra including their lunar components have been thoroughly analyzed. The lunar components seemed to be suppressed with something even below a background level. This can be possible if there are two identical processes but acting in a reversed phase. Consequently, the core constantly "follows" the Moon's position due to their mutually-interacted motion. The Moon gravitation is neutralized by Earth's core gravitation due to its displacement relative to the geometrical center of the planet. Thus, the core not only moves along the annual path mentioned above but spirals along it according to the current location of the Moon. The core position analyzed according to ENPEMF data proves its spiral motion notably that the Earth's solid core and the Moon spin in phase opposition and in different directions relative to each other during a lunar month.

The core like any other solid moving in a fluid has to produce a complicated wave system and wave wakes. A characteristic feature of ship's wakes produced by a ship sailing on the sea is a distinct topside view of wave ridges. The wavetrain makes an angle of $19.5°$ with the direction of ship riding and looks like a "moustache" diverging from the ship's head.

In our case when waves are produced by the solid core moving inside the liquid there has to appear a latitude effect, i.e. we observe a delay in wave arrivals as a measurement point gets farther from the core path. We have analyzed ENPEMF diurnal patterns recorded by MGR-01 stations spread in latitude over a territory of 700 km. This analysis has shown that time of wave arrival is constantly increasing as measurement points move northwards. Angle of wave delay is about $20°$. Thus one can assume that ENPEMF diurnal patterns recorded in middle latitudes of the Northern hemisphere not only relate to eccentric rotation of the Earth's core and lithosphere but also to waves produced in the liquid core due to its eccentric motion.

## Conclusions

In conclusion we would like to underline that when analyzing possible lithospheric sources of EM fields one should take into consideration that existing geosystems can hardly be described by a classical theory of skin-effect (Bogdanov, Pavlovich, Shuman, 2009). Therefore possibility of coming EM pulses up to the surface should not be ruled out even in case when field sources are located much deeper than the skin layer.

Let us underline one more time that strain waves produced by eccentric rotation of the Earth's core and the lithosphere are considered to be the most likely mechanism of generation of a lithospheric component of the Earth's natural pulse electromagnetic field. These waves particularly carry all the information on everything the meet along their path from the lower mantle up to the surface. Having such a mechanism of generation of spatial and temporal ENPEMF variations we do not record pulses at km-scale depths but we do record pulses originated at a skin-layer depth or less. However intensity of such near-surface pulse sources is governed by characteristics of strain waves traveling from the lower mantle up to the surface. When applying a correct methodology, thorough analysis and correct interpretation of ENPEMF data it is possible to extract much useful and necessary data including information on the core itself (Malyshkov et al., 2009a) and the strain wave propagation path.

It also should be noted that the proposed method is based on natural processes of mechanic-to-electric energy conversion in rocks. Therefore the method is able to give information on both mechanical and electrical properties of the geologic environment. Moreover it combines advantages of seismic acquisition and geoelectric survey. Pulse signals originate from lithologic and structural heterogeneities; such heterogeneities generate pulses due to rock micro movement caused by natural processes occurring in the crust. All these make the method ecologically friendly and selectively sensitive for various geologic interfaces. It is a rock-rock interface that is of great interest to specialists in searching for and exploration of any oil field. The ENPEMF method does not require any special preparation of measurement routes or blast operations; it may be carried out by few operators on foot or by using any land vehicle.

We quite understand that this new geophysical survey method is still in an initial phase of its development. For now there are no theoretical models or estimates and we have not considered the construction of subsurface cross-sections based on ENPEMF data yet; all the concepts and mechanisms of spatial and temporal ENPEMF variations require a thorough analysis and confirmation. However, if we are right in our interpretation, there may appear fundamentally new ways of investigating the Earth's structure as early as the nearest future; such fundamentally new ways and methods will excel all now-existing methods including seismic acquisition by their profoundness and information capacity. The new method can be based on recording not only the Earth's natural pulse electromagnetic field but also many other geophysical fields sensitive to subsurface strain waves produced by the Earth's core.

The authors are cordially open to cooperation and to any proposal on cooperative development of such work (E-mail: msergey@imces.ru).